\documentclass[12pt]{article}
\usepackage{psfig}
\usepackage{graphicx}

\begin{document}

\begin{center}

\bigskip

{\LARGE UHECRs deflections in the IRAS PSCz catalogue based  models  of extragalactic magnetic field} \\

{\small A. Elyiv}  \\

{\em Astronomical Observatory, Kyiv National University, Observatorna str., 3,
Kyiv 304053, Ukraine} \\
elyjiw@ukr.net \\

\end{center}

{\large \bf Abstract}\\

We present an investigation of the propagation of Ultra High Energy Cosmic Rays (UHECRs) in extragalactic magnetic field (EGMF). We use the IRAS PSCz catalogue in order to reconstruct EGMF taking into account power-law dependence between the magnetic field $B$ and the infrared luminosity density $\rho _L$, $B=K{\rho _{L}}^{\beta }$. Contrary to the previous works [11] the normalization parameters $K$ and $\beta $ have been estimated according to the observable values of magnetic field of the galaxy clusters ($B_{GC} = 0.3 - 2\, \mu$G) and assessed values of field in the voids of the Local Universe ($B_{Void} = 10^{-12} - 10^{-10}$\,G). We construct the full-sky maps of expected deflection angles of UHE protons with the arrival energies $E = 10^{20}$\,eV and $E = 4\cdot 10^{19}$\,eV in the reconstructed EGMF for two boundary cases of  "strong" ($B_{GC} = 2\,\mu$G, $B_{Void} = 10^{-10}$\,G) and "weak" ($B_{GC} = 0.3\,\mu$G, $B_{Void} = 10^{-12}$\,G) magnetic fields. It is found that average deflections of protons with observable energies above $4\cdot 10^{19}$\,eV and maximum energy in sources $E_{max} = 10^{22}$\,eV are unimportant (comparable with the errors of modern detectors) only in the case of "weak" EGMF model.

\bigskip

\section{Introduction}
The nature of UHECRs with $E>10^{19}$\,eV is one of the most challenging problems of modern astrophysics [1, 2]. At these energies the cosmic ray particles strongly interact with the cosmic microwave background and their flux should be severely attenuated according to GZK cutoff of UHECR energy spectrum [3, 4]. AGASA detector in Japan has registered 11 events with energies above $10^{20}$\,eV [5]. If galactic and extragalactic magnetic fields are not enough strong, then protons of these energies may not be significantly deflected so that some of the cosmic rays may point back to their sources. Therefore, the EGMF is of great importance for the search of UHECRs sources.

A few different methods of EGMF reconstruction are presented in literature. The authors [6, 7] described the magnetic field distribution by simple models that were connected with such large-scale structures as the Local Supercluster, the Supergalactic wall etc. Recently, several groups started developing physically more realistic models based on numerical simulations, combining the magneto-hydrodynamics of the intergalactic medium with N-body simulations of the driving gravitational dynamics of the dark matter [8, 9, 10]. The method that was presented in work [11] is realistic and relatively simple simultaneously. In this work the EGMF simulation was carried out on the basis of power-law dependence between the galaxy infrared luminosity density and strength of the magnetic field.

We use this approach in our work to reproduce the EGMF, but we realize a calibration of the field energy other way. The authors [11] set a normalization of magnetic field $B = 0.4\,\mu$G in a $1\,Mpc^{3}$ cube in the centre of the Virgo cluster and took the parameter $\beta = 2/3$ in the dependence between the infrared luminosity density and EGMF . We calibrated the simulated EGMF using observable magnetic field values in the centre of the Coma galaxy cluster and assessed values of the field in the voids of the Local Universe.

\section{Models of the extragalactic magnetic field}

The EGMF are little known theoretically and is difficult for observations. The results of [9, 12] show that the magnetic field traces baryon density. On the other hand, the Faraday rotation measurements of polarized radio sources placed within cluster of galaxies and observations of hard X-ray emission from clusters of galaxies points at connection between magnetic field and distribution of galaxies [13]. We have reconstructed the EGMF, from the hypothesis about its origin as a result of magnetized plasma injection in the extragalactic medium (within $1$\,Mpc aroung the galaxy) via supernova explosions and galactic wind action [14]. If the galactic winds  from supernova explosions magnetize the intergalactic medium in the same way as they magnetize the interstellar medium, then, under the assumption of equipartition, the magnetic field in the intergalactic medium should have the magnetic field  energy density comparable to the energy density of cosmic rays [15, 16]. The authors [11] assume that the luminosity density is proportional to the gas density. In our work we used the IRAS PSCz catalogue within $100$\,Mpc [18] for construction of EGMF model similar to [11]. We applied power-law dependence between magnetic field  $B$ and infrared luminosity density $\rho _{L}$, at a point $\overrightarrow{r}$, in the form of:
$$
B(\overrightarrow{r})=B_{0}\left(\frac{\rho _{L}(\overrightarrow{r})}{\rho _{L0}}\right)^\beta ,
$$
where $\rho _{L0}$ and $B_{0}$  is the values of infrared luminosity density and magnetic field, and $\beta$  is the free parameter. 
For calibration of this equation we took the conditions in the Coma cluster centre ($l = 58^{\circ}$, $b = 88^{\circ}$, $r = 96$ Mpc, $\rho _{L0} = 1.1\cdot 10^{10} L_{\odot}/{Mpc}^{3}$) and in the voids of the Local Universe. By comparing the Coma radio synchrotron spectrum with excess radiation, interpreted as the inverse Compton scattering of cosmic microwave background photons by same electron population, volume-averaged magnetic field of $B \sim 0.1\, \mu$G have been deduced in [20], $0.2\,\mu$G in [21], and $0.4\,\mu$G in [22]. Faraday rotation measurements gave $B\sim 2\mu$\,G [17], $8.5\,\mu$G [19]. The obtained values of magnetic field may be divided into  two groups: $B<1 \mu G$ and $B>1 \mu G$. We took the value $B_{0}=0.3\,\mu$G as representative for the first group, and $B_{0}=2.0\,\mu$G for the second group.

The data about Local Universe voids were taken from [25]. The mean value of void infrared luminosity density $\rho_{Void}$ is equal to $9.9\cdot 10^{6} L_{\odot}/Mpc^{3}$. We took the value $B_{Void} = 10^{-10}$\,G as representative for the upper limit of EGMF in voids and the lower limit $B_{Void} = 10^{-12}$\,G for the magnetic fields in the voids from [24]. Taking into account the data described above the two approximate dependences between magnetic field  and infrared luminosity density were obtained: the model of "weak" field ($B_{0}=0.3\,\mu$\,G, $B_{Void}=10^{-12}$\,G) --
$$
B(\overrightarrow{r})=0.3\left(\frac{\rho _{L}(\overrightarrow{r})}{1.1\cdot 10^{10} L_{\odot}/{Mpc}^{3}}\right)^{1.8} \mu G,
$$
and the model of "strong" field ($B_{0}=2.0\,\mu$G, $B_{Void}=10^{-10}$\,G) - 
$$
B(\overrightarrow{r})=2.0\left(\frac{\rho _{L}(\overrightarrow{r})}{1.1\cdot 10^{10} L_{\odot}/{Mpc}^{3}}\right)^{1.4} \mu G,
$$
The magnetic field distribution  along the three fiducial lines through the Virgo cluster, the Perseus cluster and the Coma cluster are shown in the Fig. 1. To avoid zero magnetic field value we assigned the minimal EGMF ($B_{Void}$) to volumes with $\rho _{L}(\overrightarrow{r}) = 0$.

\begin{figure}
\centerline{\includegraphics[angle=0, width=17cm]{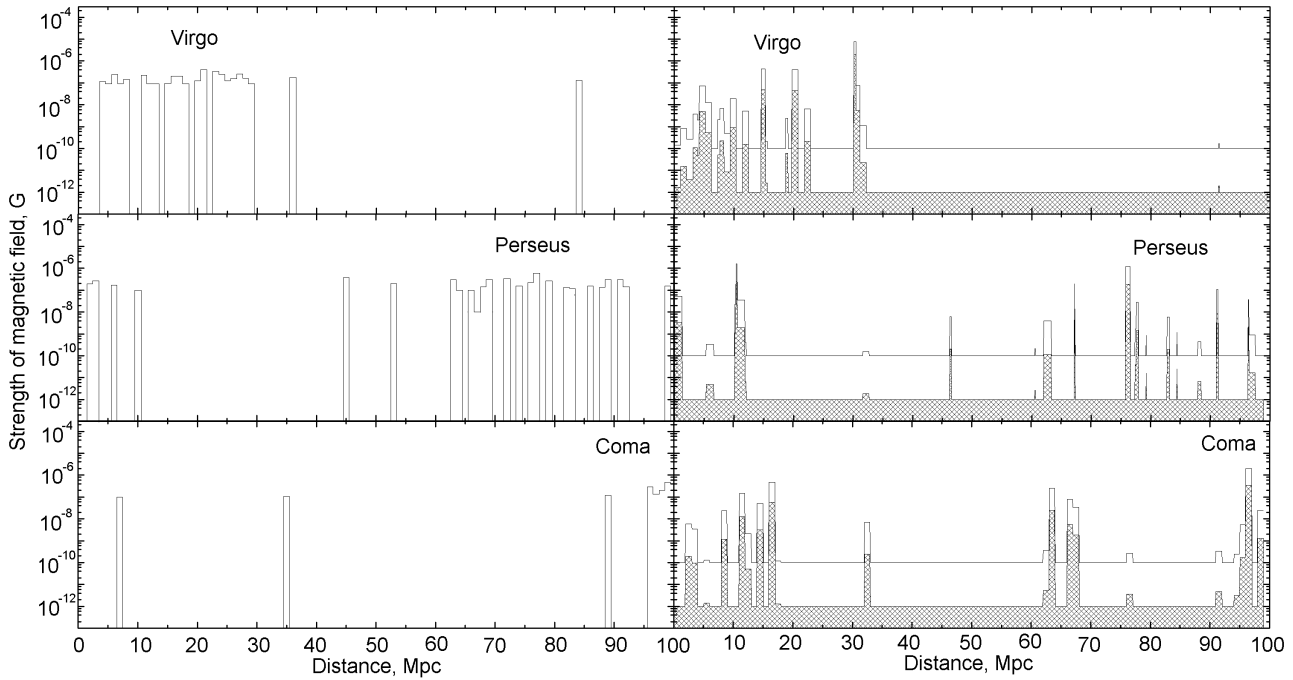}}
\caption{The magnetic field distribution  along the three fiducial lines through the Virgo cluster, the Perseus cluster and the Coma cluster from [11] (left panel) and from our simulation (right panel). The shaded and unshaded regions on the right panel represent the models of  the "weak" and "strong" magnetic fields, respectively.}
\end{figure}
It is reasonable that direction of the EGMF induction is random. Coherent lengths of EGMF $l_{c}$ in the galaxy cluster regions lie in the range from $10$ to $100 kpc$ [26, 27], and are approximately equal to $1$ Mpc for intergalactic medium with $B<10^{-9}$\,G [26]. We have used the following conditions: if $B<10^{-9}$\,G then value of $l_{c}$ is equal to $1$\,Mpc, if $B > 10^{-9}$\,G then $l_{c}$ took on the values $250$, $50$ and $10$\,kpc.

\section{The deflection of UHECRs in the extragalactic magnetic field}

We have studied the UHECR deflections in the cases of different EGMF models. For this we injected antiprotons from the Earth and followed their trajectory taking into account energy losses of the UHECRs due to the interactions with cosmic microwave background photons [23].\\

\begin{figure}[ht]
\centerline{\includegraphics[angle=0, width=14cm]{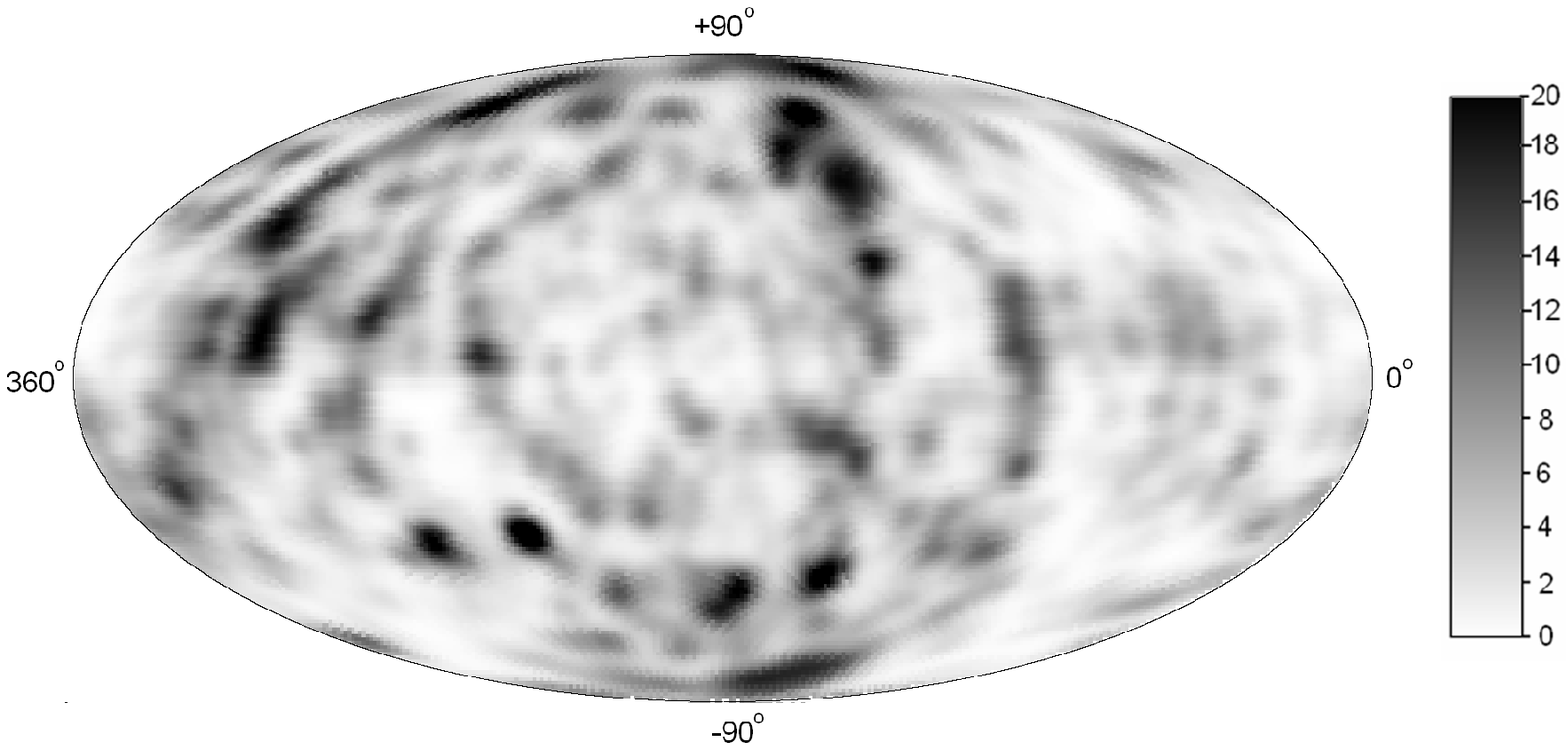}}
\,
\centerline{\includegraphics[angle=0, width=14cm]{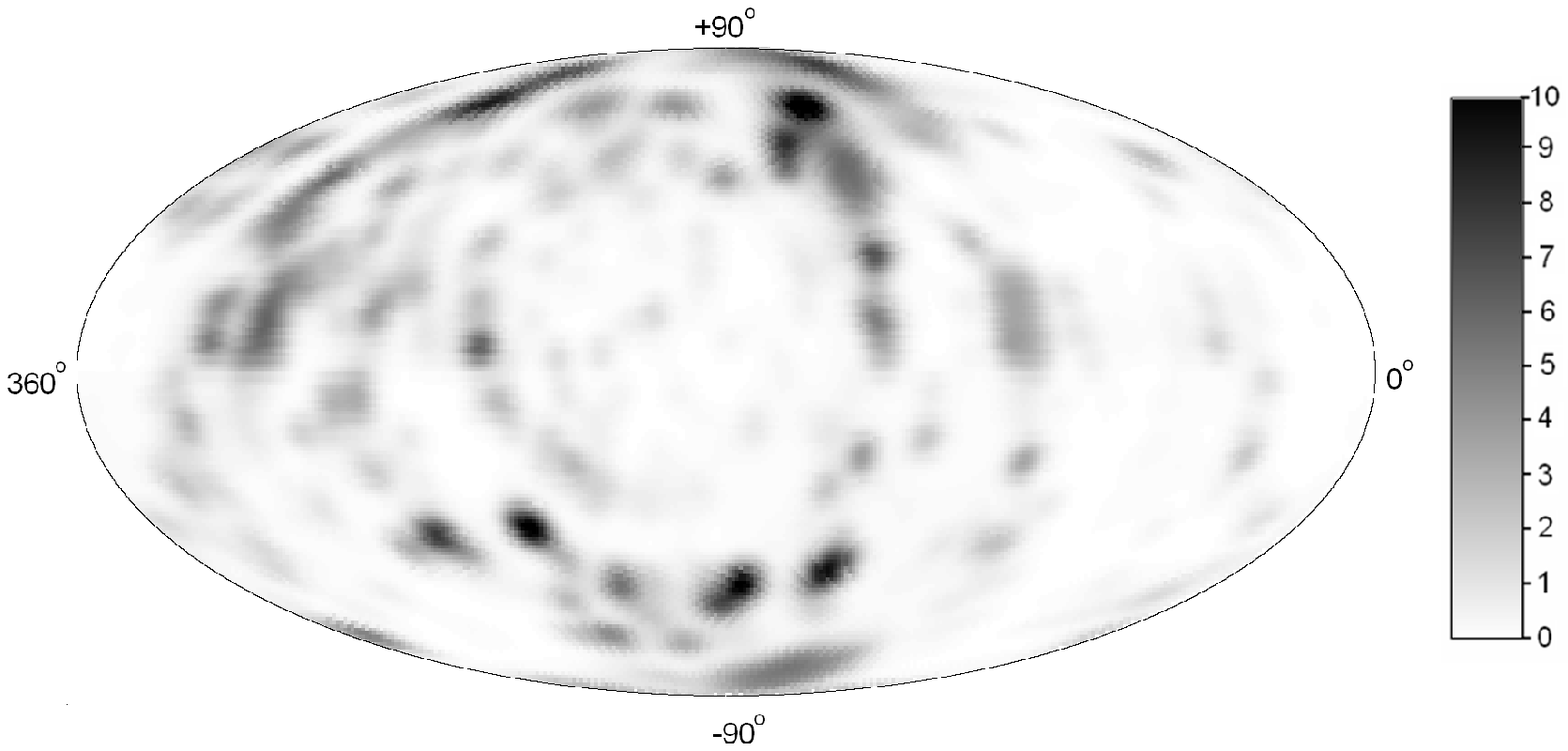}}
\caption{The deflection angles of UHE protons with arrival energy $E = 4{\cdot} 10^{19}$\,eV (top panel) and $E = 10^{20}$\,eV (bottom panel) through propagation in "weak" model of extragalactic magnetic field with $l_{c} = 50$\,kpc. Both maps are smoothed with Gaussian kernel of $3^{\circ}$. The coordinate system is galactic. Deflections in degrees.}
\end{figure}

Table 1: The average deflection of the UHECRs in the different EGMF models\\

\begin{center}
\begin{tabular}{|c|c|c|c|c|c|}
\hline
 &  & \multicolumn{4}{|c|}{Energy, eV}\\ 
\hline
$l_{c}$, kpc & Model & $4\cdot 10^{19}$ & $6\cdot 10^{19}$ & $10^{20}$ & $3\cdot 10^{20}$\\
\hline
250 & weak & $9.5^{\circ}$ & $7.1^{\circ}$ & $2.5^{\circ}$ & $0.4^{\circ}$\\
\cline{2-6}
 & strong & $23^{\circ}$ & $19^{\circ}$ & $11^{\circ}$ & $2.9^{\circ}$\\
\hline
50 & weak & $5.3^{\circ}$ & $3.4^{\circ}$ & $1.2^{\circ}$ & $0.2^{\circ}$\\
\cline{2-6}
 & strong & $18^{\circ}$ & $14^{\circ}$ & $7^{\circ}$ & $1.3^{\circ}$\\
\hline
10 & weak & $2.4^{\circ}$ & $1.5^{\circ}$ & $0.5^{\circ}$ & $0.1^{\circ}$\\
\cline{2-6}
 & strong & $12^{\circ}$ & $9.2^{\circ}$ & $3.3^{\circ}$ & $0.6^{\circ}$\\
\hline
\end{tabular}
\end{center}

\newpage
The trajectories were followed until  the  distances of particles from the Galaxy reached $100$\,Mpc or their energies reached $10^{22}$\,eV. 
The average deflections of the UHECRs significantly depend on the coherent length value $l_{c}$ and type of EGMF model (Tab. 1). So, the average deflection for the "strong" model is 3 -- 6 times as much as that for the "weak" model. The deflection maps of protons in the case of the "weak" magnetic field model for $l_{c}=50$\,kpc are presented in the Fig. 2. As can be seen from these maps, regions with significant deflections (above $15^{\circ}$ for UHECRs with arrival energy $4\cdot 10^{19}$\,eV and above $7^{\circ}$ for $10^{20}$\,eV) correspond to locations of main galaxy clusters and near galaxies with high infrared luminosities. Although the most part of the sky area is occupied by regions with deflection values less then $4^{\circ}$ for $E = 4\cdot 10^{19}$\,eV and $1^{\circ}$ for $E = 10^{20}$\,eV. Thus it is appropriate to search of correlation between the UHECRs (with arrival energy above $4\cdot 10^{19}$\,eV) arrival directions and their possible sources in the case of "weak" model.

\section{Conclusion}

In this work we have investigated the influence of EGMF on the UHECRs propagation. For this purpose we have reconstructed EGMF taking into account power-law dependence between magnetic field and infrared luminosity density of galaxies in the Local Universe. As distinct from [11] this dependence was normalized to the observable values of the magnetic field in the galaxy clusters and the assessed values of the field in the voids of the Local Universe. We have found the value of exponent $\beta$ in  power-low dependence  for the two boundary cases of magnetic fields in the Coma cluster and in the voids. For the "weak" EGMF model ($B_{GC} = 0.3 \mu$\,G, $B_{Void}=10^{-12}$\,G) the parameter $\beta$ is equal to 1.8 and for the "strong" field model ($B_{GC}=2.0 \mu$\,G, $B_{Void}=10^{-10}$\,G) $\beta = 1.4$. The average values of UHECR deflection depend on the type of EGMF model. It was shown, that deflection angles for protons with the arrival energy equal to $4\cdot 10^{19}$\,eV significantly exceed errors of modern detectors in case of "strong" EGMF model. For "weak" EGMF model the deflections lie mainly in the range of $1^{\circ} - 5^{\circ}$, therefore charged particle astronomy should be possible in that case. 

On the basis of the average deflection comparison we can conclude, that our "weak" model of EGMF agrees with the model that is presented in [11]. On the other hand our results differ from [9], where the authors constructed EGMF model using simulations of large-scale structure formation to study the build-up of magnetic fields in the intergalactic medium. In average our deflection angles are larger then obtained in [9]. Our results are in good agreement with [10, 12] where the significant UHECR deflections in simulated EGMF were  found. 

\bigskip

{\bf ACKNOWLEDGMENTS.} I am very grateful to Dr. Bohdan Hnatyk for the permanent attention to my work and useful discussion. 
{}
\end{document}